\documentstyle[12pt]{article}
\begin{document}
\begin{center}
{\bf ALTERNATE ROUTE TO SOLITON SOLUTIONS\\
IN HYDROGEN--BONDED CHAINS}
\end{center}

\bigskip

\begin{center}
D. Bazeia, J. R. S. Nascimento, and D. Toledo
\end{center}

\begin{center}
Departamento de F\'\i sica, Universidade Federal da 
Para\'\i ba\\
Caixa Postal 5008, 58051-970 Jo\~ao Pessoa, Para\'\i ba, Brazil
\end{center}

\vskip 3cm

\begin{center}
Abstract
\end{center}

In this paper we offer an alternate route for investigating soliton solutions in hydrogen-bonded chains. This is done by
examining a class of systems of two coupled real scalar fields. We show that this route allows investigating several models for hydrogen-bonded chains in a unified manner. We also show how to investigate interesting issues, in particular the one concerning classical or linear stability of solitonic solutions.

\newpage

\section{Introduction}

Hydrogen-bonded materials that appear in condensed matter, and in organic and biological systems, may 
transport energy and charge, and this transport
seems to be due to proton transfer along the
hydrogen-bonded chains\cite{gla75,ntn83,dni83,pne88,ptz89}. The conduction via proton migration along the chain
in hydrogen-bonded materials appears as follow. We can 
think of a hydrogen-bonded material as being genericaly represented by
$$
{\cdots}{\rm{X-H}}
{\cdots}{\rm{X-H}}{\cdots}
{\rm{X-H}}{\cdots}
$$
where $(-)$ and $(\cdots)$ stand for the covalent and the hydrogen bond, respectively. Here ${\rm X}$ represents a group of atoms - for ice it is OH, where O stands for oxygen. The hydrogen bond bridges the system in many repetitions of the unit cell $ {\rm{X-H}}{\cdots}$, and the system is usually considered as a unidimensional macroscopic chain.
 
In such a macroscopic chain, protons are transfered via two distinct mechanisms. The intrabond mechanism accounts for proton migration through the bridge, and this proton migration changes the role of the covalent $(-)$ and the hydrogen bond $(\cdots)$. In this case there is a local disturbance of the neutral charge of the chain, and so this defect is called an ionic defect. The interbond mechanism accounts for the interchange of position of the covalent bond. Here there is a local orientational change of the covalent bond, and so this defect is called an orientational (or Bjerrum) defect. 

Hydrogen-bonded materials have been studied by many authors. 
A particularly interesting investigation of soliton 
solutions for hydrogen-bonded chains was introduced in Ref.~{\cite{psz91}}. In this case, the longitudinal dynamics of protons was investigated within the context of
one-component models for hydrogen-bonded chains.

In a more recent work a specific model of hydrogen-bonded chain was examined \cite{xzh96}.  Here the model is of the two-component type, and is based on the assumption that the coupling between the proton sublattice (the first component) and the acoustic mode of the heavy-ion sublattice (the second component) is linear. In this case the solitonic solution is a kink or a bell-shape soliton, depending of the specific parameters which control the proton on-site potential, which is represented by a symmetric (kink) or asymmetric (bell-shape) double-well potential.

In the present paper, we get inspiration from recent works \cite{bds95,bsa96,brs96} to offer an alternate route for examining soliton solutions in hydrogen-bonded chains. This will be done by introducing a class of systems of two coupled scalar fields. Here, however, we shall work with the same assumption considered in \cite{xzh96}, which states that the coupling between the proton sublattice and the acoustic mode of the heavy-ion sublattice is linear. 

As we are going to show, the class of systems we shall introduce can be used to map the model investigated in \cite{xzh96} and other models, in a unified manner. Within this context, our investigations provide a new route for examining soliton solutions in hydrogen-bonded materials. Before doing this, however, let us first introduce the class of systems of two coupled fields. 

\section{Two coupled real scalar fields}
\label{sec:systems}

The investiagtions we are now going to do is inspired on former works \cite{bds95,bsa96,brs96}, from where we borrow notation, which is standard notation \cite{raj82} in Field Theory. The class of systems we shall investigate is described by the following Lagrange density
\begin{equation}
\label{eq:model}
{\cal L}=\frac{1}{2}(1+a^2)
\partial_{\alpha}\phi\partial^{\alpha}\phi+
\frac{1}{2}b^2\partial_{\alpha}\chi
\partial^{\alpha}\chi+ab\partial_{\alpha}\phi
\partial^{\alpha}\chi - U(\phi).
\end{equation}
We are working in $(1+1)$ dimensions. Hence $\alpha=0,1$,
and $x^{\alpha}=(x^0=t,x^1=x)$, and $x_{\alpha}=(x_0=t,x_1=-x)$.
Here $\phi$ and $\chi$ are real scalar fields, dimensionless, and $a$ and $b$ are real and dimensionless parameters. The potential $U(\phi)$ is chosen as
\begin{equation}
\label{eq:pot}
U(\phi)=\frac{1}{2}\left(\frac{d H}{d\phi}\right)^2,
\end{equation}
and is defined from the function $H=H(\phi)$, which is a smooth but otherwise arbitrary function of the field $\phi$. The reason for the above choice $(\ref{eq:pot})$ for the potential will appear later.

In the above model, evidently, each function $H(\phi)$ defines a specific system, and so we have a class of systems of two coupled real scalar fields. The coupling
between the two fields is a derivative coupling -- the third term in $(\ref{eq:model})$ -- and is the same for every specific system one introduces by some specific function $H(\phi)$. We then recognize that systems belonging to the above class are different when they differ on the 
self-interaction term, or better, on the potential for the scalar field $\phi$. 

The Euler-Lagrange equations that follow from the above 
system can be written as 
$$
\partial_{\alpha}\partial^{\alpha}\phi+
H_{\phi}H_{\phi\phi}+
a^2\partial_{\alpha}\partial^{\alpha}\phi+
ab\partial_{\alpha}\partial^{\alpha}\chi=0,
$$
and
$$
b^2\partial_{\alpha}\partial^{\alpha}\chi+
ab\partial_{\alpha}\partial^{\alpha}\phi=0.
$$
These equations can be rewriten in the following form
\begin{equation}
\label{eq:emphi}
\partial_{\alpha}\partial^{\alpha}\phi+
H_{\phi}H_{\phi\phi}=0,
\end{equation}
and
\begin{equation}
\label{eq:emchi}
a\partial_{\alpha}\partial^{\alpha}\phi+
b\partial_{\alpha}\partial^{\alpha}\chi=0.
\end{equation}
For static field configurations they become
\begin{equation}
\label{eq:emphis}
\frac{d^2\phi}{dx^2}=H_{\phi}H_{\phi\phi},
\end{equation}
and
\begin{equation}
\label{eq:emchis}
a\frac{d^2\phi}{dx^2}+
b\frac{d^2\chi}{dx^2}=0.
\end{equation}
According to the standard route to solitons \cite{raj82}, the above equations of motion are the equations we have to solve to find soliton solutions.

Before following this route, however, let us investigate the energy corresponding to static field configurations. Here we use the Lagrange density $(\ref{eq:model})$ to write the energy as $E=E_{B}+E'$, where $E_B$ is
\begin{equation}
\label{eq:energy}
E_{B}=H[\phi(\infty)]-
H[\phi(-\infty)],
\end{equation}
and $E'$ has the form
\begin{equation}
E'=\frac{1}{2} \int^{\infty}_{-\infty}dx \Biggl{[}
\left(\frac{d\phi}{dx} - H_{\phi}\right)^2+
\left(a\frac{d\phi}{dx}+b\frac{d\chi}{dx}\right)^2+
2U-\left(\frac{dH}{d\phi}\right)^2
\Biggr{]}.
\end{equation}
Here we recognize that the choice $(\ref{eq:pot})$ for the potential lead us to the result that the energy corresponding to static field configurations can get to its lower bound, $E_B$, and this happens when one sets
\begin{equation}
\label{eq:foeqphi}
\frac{d\phi}{dx}=\frac{dH}{d\phi},
\end{equation}
and
\begin{equation}
\label{eq:foeqchi}
a\frac{d\phi}{dx}+ b\frac{d\chi}{dx}=0.
\end{equation}
As we can check straightforwardly, the above first-order equations $(\ref{eq:foeqphi})$ and $(\ref{eq:foeqchi})$ solve the equations of motion $(\ref{eq:emphis})$ and $(\ref{eq:emchis})$. These results are interesting since we can search for soliton solutions in coupled nonlinear (but now) first-order differential equations, and we already know that such solitons present minimum energy. 

Here we notice that $\chi(x)={\bar{\chi}}-(a/b)\phi(x)$, with ${\bar{\chi}}$ constant, is the choice that solves the first-order equation $(\ref{eq:foeqchi})$. This is the solution we use for the $\chi$ field. Furthermore, since our system is Lorentz-invariant, we can get time-dependent solutions by just boosting the static solutions. Hence, in order to mantain Lorentz invariance we must assume that $\chi(x,t)={\bar{\chi}}-(a/b)\phi(x,t)$. In this case both
$\phi(x,t)$ and $\chi(x,t)$ satisfy the time-dependent equation of motion $(\ref{eq:emchi})$. This is an interesting result, because now we can investigate classical or linear stability of the soliton solutions by just examining the equation for the $\phi$ field. 

To make this investigation explicit, let us now write
\begin{equation}
\phi(x,t)=\phi(x)+\eta(x,t),
\end{equation} 
where $\eta(x,t)$ is the fluctuation about the classical solution $\phi(x)$. We substitute the above $\phi(x,t)$ 
into the time-dependent equation of motion $(\ref{eq:emphi})$ to get, up to first-order in $\eta$,
\begin{equation}
-\frac{\partial^2\eta}{\partial x^2}+
(H^2_{\phi\phi}+ H_{\phi}H_{\phi\phi\phi})\,\eta=
-\frac{\partial^2\eta}{\partial t^2}.
\end{equation}
In this case we recall that the choice $\chi(x,t)= {\bar{\chi}}-(a/b)\phi(x,t)$ is fully compatible with the investigation we are now dealing with. Furthermore, by assumption the classical solution $\phi(x)$ is static, and so the term $H^2_{\phi\phi}+ H_{\phi}H_{\phi\phi\phi}$ in the above equation is time-independent. Hence we can write $\eta(x,t)=\eta(x)T(t)$, in order to separate space and time in the fluctuation. For this reason we can investigate classical or linear stability \cite{bsa96} by examining the following time-independent Schr\"odinger operator
\begin{equation}
S_2=-\frac{d^2}{dx^2}+
H^2_{\phi\phi}+H_{\phi}H_{\phi\phi\phi}.
\end{equation}

On the other hand, we can use the first-order equation $(\ref{eq:foeqphi})$ to introduce the following operators
\begin{equation}
S_1^{\pm}=\pm\frac{d}{dx}+H_{\phi\phi}.
\end{equation}
It is not hard to see that these first-order operators are adjoint of each other. Furthermore, they can be used to write the second-order Schr\"odinger operator as $S_2=S_1^{+}S_1^{-}$. These facts ensure that $S_2$ is positive semi-definite, and hence the static field $\phi(x)$ that solves the first-order equation $(\ref{eq:foeqphi})$ is classically or linearly stable.

Before ending this section, let us now briefly comment on some topological issues. Since we are working with a bidimensional spacetime, we can introduce the following vector
\begin{equation}
\label{eq:current}
j^{\alpha}=\epsilon^{\alpha\beta}\partial_{\beta} H(\phi),
\end{equation}
where $\epsilon^{\alpha\beta}$ is the Levi--Civita tensor, which obeys $\epsilon^{00}=\epsilon^{11}=0$ and $\epsilon^{01}=-\epsilon^{10}=1$. Hence we notice that
$\partial_{\alpha}j^{\alpha}=0$, and so there is a conserved quantity, the so-called topological charge, which can be written in the following form
\begin{equation}
\label{eq:topcharge}
Q_T =\int_{-\infty}^{\infty} dx \, j^0=
H[\phi(\infty)]-H[\phi(-\infty)].
\end{equation}
Here we notice that the current density $(\ref{eq:current})$ 
is defined \cite{brs96} in terms of $H=H(\phi)$, in order to make the topological charge equal to the energy of the static field, as given in $(\ref{eq:topcharge})$. Evidently, other definitions for the current density can be used, but they make sense if and only if the potential has two or more absolute minima \cite{raj82}. This is so because the field
configuration has to get to the absolute minima of the potential, asymptotically, to make the energy finite.
Within this context, the definition we are using is very natural, because it uses $H(\phi)$, and $H(\phi)$ is just the function that defines the potential. For potentials that have only one absolute minimum, nontrivial field configurations must go to that value, asymptotically. Such solutions can not have topological charge, and this is the reason to name them
nontopological solutions.

\section{Two-component hydrogen-bonded chains}
\label{sec:models}

In this section we are concerned with mapping the class of systems introduced in the former section to hydrogen-bonded chains. Toward this goal, let us first recognize that when we set $a=b=0$, the above systems can be used to map the continuum version of one-component models for hydrogen-bonded chains. In this case the real scalar field $\phi$ has to be identified with the proton degree of freedom in the unidimensional chain.

To model two-component systems, we can naturaly use the scalar field $\phi$ to map the proton degree of freedom, in the proton sublattice. Within this context, we now take the scalar field $\chi$ to map the acoustic mode of the heavy-ion sublattice. Hence we notice that our class of systems can describe hydrogen-bonded systems with different on-site proton potentials, but with qualitatively the same linear coupling between the proton sublattice and the acoustic mode of the heavy-ion sublattice. As we have already shown, such a coupling leads to the solution $\chi(x,t)={\bar{\chi}}- (b/a)\,\phi(x,t)$, where $\phi(x,t)$ is obtained by boosting
the static field $\phi(x)$ that solves the first-order equation $(\ref{eq:foeqphi})$, and hence the static second-order equation of motion. 

In order to illustrate the procedure, let us consider some explicit examples. Firstly we introduce
\begin{equation}
\label{eq:sinegordon}
H_{1\phi}=\lambda\sin(\phi),
\end{equation}
where $\lambda$ is a real parameter, with dimension of energy. Here we get
\begin{equation}
H_1=-\lambda\cos(\phi),
\end{equation}
and so the potential has the form
\begin{equation}
U_1(\phi)=\frac{1}{2}\lambda^2\sin^2(\phi).
\end{equation}
This is the sine-Gordon system, and the first-order 
equation $(\ref{eq:foeqphi})$ becomes
\begin{equation}
\frac{d\phi}{dx}=\lambda\sin(\phi),
\end{equation}
and presents, for instance, the following soliton solutions
\begin{equation}
\phi(x)=\pm 2\arctan\exp(\lambda x).
\end{equation}
Here we notice that sign$(\lambda)$, which is unimportant to specify the system, gives the corresponding soliton and antisoliton solutions. See that the above solutions connect
pairs of adjacent minima in the set of minima energy states $\phi=0$ and $\phi_{\pm}=\pm\pi$ of the system. See also that there exist an infinity set of solutions, which connect pairs of adjacent minima in the infinity set of minima energy states. As one knows, the sine-Gordon system is very rich, 
and there exist more solutions.

As another system, let us choose
\begin{equation}
\label{eq:phi4}
H_{2\phi}=\lambda(v^2-\phi^2),
\end{equation}
where $v$ is real and positive. In this case we have
\begin{equation}
H_2=\lambda\phi\left(v^2-\frac{1}{3}\phi^2\right),
\end{equation}
and the potential is given by
\begin{equation}
U_2(\phi)=\frac{1}{2}\lambda^2(\phi^2-v^2)^2.
\end{equation}
This is the $\phi^4$ system, and the first-order equation 
$(\ref{eq:foeqphi})$ becomes
\begin{equation}
\frac{d\phi}{dx}=\lambda(v^2-\phi^2),
\end{equation}
which presents the kink solution
\begin{equation}
\phi(x)=v\tanh(\lambda v x).
\end{equation}
Here we also notice that sign$(\lambda)$, which is unimportant to specify the system, gives the kink $(\lambda>0)$ and antikink $(\lambda<0)$ solutions. See that these solutions
connect the two minima energy states $(\phi_{\pm}=\pm v)$ of this system.

As yet another system, let us choose
\begin{equation}
\label{eq:phi6}
H_{3\phi}=\lambda\phi(v^2-\phi^2).
\end{equation}
In this case we have
\begin{equation}
H_3=\frac{1}{2}\lambda\phi^2
\left(v^2-\frac{1}{2}\phi^2\right),
\end{equation}
and the potential is given by
\begin{equation}
U_3(\phi)=\frac{1}{2}\lambda^2\phi^2(\phi^2-v^2)^2.
\end{equation}
This is the $\phi^6$ system, and the first-order equation 
$(\ref{eq:foeqphi})$ becomes
\begin{equation}
\frac{d\phi}{dx}=\lambda\phi(v^2-\phi^2),
\end{equation}
It presents the kink solutions
\begin{equation}
\phi_{\pm}(x)=\pm\left\{ \frac{1}{2}v^2
[1\pm\tanh(\lambda v^2 x)]\right\}^{1/2}.
\end{equation}
Here we notice once again that sign$(\lambda)$, which is unimportant to specify the system, also gives all the solutions. See that these solutions connect pairs of adjacent minima in the set of three minima energy states $(\phi=0, \phi_{\pm}=\pm v)$ of this system.

The above two first examples are standard models for the on-site potential in hydrogen-bonded chains -- see, for instance, Ref.~{\cite{psz91}. In particular, in \cite{psz91}
a more general potential was also considered. Here we refer to the following choice
\begin{equation}
H_{4\phi}=\frac{\lambda}{1-d_1^2}\,
\frac{\cos(\phi)-d_1}{1-d_2[\cos(\phi)-d_1]},
\end{equation}
where $d_1,d_2$ are real and dimensionless parameters, with $0\le d_1 < 1$. 
We notice that when $d_1=d_2=0$, we get back to the sine-Gordon model already investigated. Yet, for $d_2=0$ we get to the double sine-Gordon model.

Despite the complexity of $H(\phi)$, we can state on general grounds that for systems bellonging to the above class we have to deal with a first-order equation, instead of the usual second-order equation of motion. As we have already shown, we know that the soliton solutions are stable and have minimum energy. The energy is obtained from $(\ref{eq:energy})$, and so we just need $H(\phi)$ and the asymptotic values of the field. After recalling that a topological soliton connects two adjacent minima of the potential, we immediately get the asymptotic values of the field, from which we write the energy easely. As an illustration, the above $H_{4\phi}$ can be
integrated to give $H_{4}$, which is an important quantity. In the one-component model of Ref.~{\cite{psz91}} this $H_{4}$ plays the role of the general function $A_{K}$ there introduced in Sect.~{II.E}. Here, however, we have shown explicitly that the solitonic solutions have minimum energy, and are classically stable.

The model $(\ref{eq:phi6})$ with the $\phi^6$ potential is a
new example. In this case there are the symmetric minimum $(\phi=0)$, and two asymmetric minima $(\phi_{\pm}=\pm v)$, and so this system seems to be richer then the $\phi^4$ system, in which we have only two asymmetric minima,
namely $(\phi_{\pm}=\pm v)$. We have added this example to further illustrate the general procedure.

\section{Other two-component models}

The class of systems we have being investigating is
specifyed by the potential $(\ref{eq:pot})$. However, 
we may find reasons to abandon this class of systems to consider other models. Such a situation has occurred, for instance, in Ref.~{\cite{xzh96}}, where an asymmetric
$\phi^4$ potential is used to find bell-shape soliton solutions. This is the case we want to consider in the following. Before doing this, however, without loosing generality let us first shift the $\phi$ field in the 
$\phi^4$ potential given in the former section. Here we 
change $\phi\to\phi\pm v$ to get
\begin{equation}
\label{eq:potenphi4}
{\bar U}(\phi)=2\lambda^2 v^2\phi^2\pm 2\lambda^2 v\phi^3+
\frac{1}{2}\lambda^2\phi^4.
\end{equation}
Of course, this is still the potential for kink solitons we examined in the former section, shifted by the value we have just introduced. Here the solitonic solutions are given by
\begin{equation}
\phi_{\pm}(x)=\mp v +v\tanh(\lambda v x).
\end{equation}

To get to the case of an asymmetric $\phi^4$ potential, let us now add to the above potential $(\ref{eq:potenphi4})$ the following contribution
\begin{equation}
U'(\phi)=-2\epsilon\lambda^2 v^2\phi^2,
\end{equation}
Here $\epsilon$ is real, positive, and dimensionless. In this case we obtain
\begin{equation}
\label{eq:potnontopo}
U(\phi)=2(1-\epsilon)\lambda^2 v^2\phi^2\pm 2\lambda^2 v\phi^3+\frac{1}{2}\lambda^2\phi^4,
\end{equation}
and so it does not bellong to the class of systems we have introduced in Sect.~{\ref{sec:systems}}. To search for solutions we have to deal with the equation of motion, which is, for static configuration,
\begin{equation}
\frac{d^2\phi}{dx^2}=4(1-\epsilon)\lambda^2 v^2\phi\pm 6\lambda^2 v\phi^2+2\lambda^2\phi^3.
\end{equation}
This equation can be integrated to give
\begin{equation}
\phi_{\pm}(x)=\pm\frac{2v(1-\epsilon)}{1+\sqrt{\epsilon}\,
\cosh(2\lambda v\sqrt{1-\epsilon}\, x)}.
\end{equation}
Here we notice that the above bell-shape solitons are nontopological solutions, and so we do not have any good reason to believe on their stability -- recall that the proof of classical stability already introduced does not work in this case. For this reason, stability of the above bell-shape 
soliton solutions has to be examined independently, but this is out of the scope of the present paper. We shall return to the issue concerning the presence of nontopological solitons in systems of coupled scalar fields in another paper.

\section{Ending comments}

In this paper we have shown explicitly that the class of systems introduced in Sect.~{\ref{sec:systems}} can be 
used to model hydrogen-bonded chains. As we have also shown 
in Sect.~{\ref{sec:models}}, this form of modelling two-component systems of hydrogen-bonded materials is specific, and works if and only if the two components couple linearly, in the form of a derivative coupling. 

The most interesting results we have found in the present route to soliton solutions in hydrogen-bonded chains can be summarized as follows. To search for solitonic solutions, it sufices to deal with first-order differential equations, which solve the corresponding equations of motion for static configurations. The solitonic solutions one may find present minimum energy and are classically or linearly stable. In addition, we have also introduced a way of accounting for topological properties of solitonic solutions.

The above results encourage us to search for more general systems of coupled scalar fields, with the aim of perhaps finding classes of systems that lead to a unified way of investigating two-component models of hydrogen-bonded chains, without assuming linear coupling between the two components from the beginning. This and other related issues are presently under consideration.

\bigskip

We would like to thank R. F. Ribeiro for some interesting discussions. DT is grateful to Conselho Nacional de Desenvolvimento Cient\'\i fico e Tecnol\'ogico, CNPq,
Brazil, for a fellowship.

\end{document}